\documentclass[conference]{IEEEtran}

\usepackage[cmex10]{amsmath}
\usepackage[dvipdfmx]{graphicx}
\usepackage{array}
\usepackage{mdwmath}
\usepackage{mdwtab}
\usepackage{amssymb}
\usepackage{color}

\newtheorem{theorem}{Theorem}
\newtheorem{lemma}{Lemma}

\newtheorem{remark}{Remark}
\newtheorem{proposition}{Proposition}
\newtheorem{example}{Example}
\newcommand{\qed}{\hfill \IEEEQED}

\DeclareMathAlphabet{\bm}{OML}{cmm}{b}{it}

\newcommand{\rom}[1]{\mathrm{#1}}
\newcommand{\san}[1]{\mathsf{#1}}

\allowdisplaybreaks[2]

\begin{document}
%
\title{Strong Converse and Second-Order Asymptotics of Channel Resolvability 
}



%
\author{\IEEEauthorblockN{Shun Watanabe\IEEEauthorrefmark{1} and
Masahito Hayashi\IEEEauthorrefmark{2}}
\IEEEauthorblockA{\IEEEauthorrefmark{1}Department of  Information Science and Intelligent Systems, 
University of Tokushima, Japan, \\
and Institute for System Research, University of Maryland, College Park. \\
Email: shun-wata@is.tokushima-u.ac.jp \\
\IEEEauthorrefmark{2}Graduate School of Mathematics, Nagoya University, Japan, \\
and Centre for Quantum Technologies, National University of Singapore, Singapore. \\
E-mail: masahito@math.nagoya-u.ac.jp}}


\maketitle

\begin{abstract}
We study the problem of channel resolvability for fixed i.i.d.~input distributions and
discrete memoryless channels (DMCs), and derive the strong converse theorem for 
any DMCs that are not necessarily full rank. We also derive the optimal second-order
rate under a condition. Furthermore, under the condition that
a DMC has the unique capacity achieving input distribution, 
we derive the optimal second-order rate of 
channel resolvability for the worst input distribution.
\end{abstract}


%
\IEEEpeerreviewmaketitle

\section{Introduction}

We study the problem of channel resolvability introduced by Han-Verd\'u \cite{han:93}
(see also \cite[Sec.~6.2]{wyner:75b}). In addition to theoretical interest as a random number
generation problem, channel resolvability has a lot of applications in problems of information theory.
First, channel resolvability can be used to show the converse coding theorem for identification via channels,
and this direction of research has been extensively studied by many researchers \cite{han:93,steinberg:98,ahlswede:02,hayashi:06b,oohama:13}.
Second, channel resolvability can be used as a building block of wiretap channel codes \cite{csiszar:96,cai:04,devetak:05,hayashi:06b,bloch:11b}.
Third, channel resolvability can be used as a building block of channel simulation, 
which in turn can be used as a building block of certain coding problems (eg.~\cite{bennett:02,winter:02,luo:09,datta:13, cuff:12,watanabe:13e}).

Despite its importance, our understanding of channel resolvability is far from complete
even for discrete memoryless channels (DMCs). 
For instance, the optimal rate of channel resolvability for fixed i.i.d.~input distribution $p$ 
is not known. In \cite{han:93}, Han-Verd\'u showed it is less than or equal to the mutual information $I(p,W)$,
and they also showed an example such that this bound is {\em not} tight \cite[Example 1]{han:93}. In \cite{han:93b}, Han-Verd\'u
showed that $I(p,W)$ is indeed the optimal rate for the class of channels called {\em full rank}.
In this paper, we derive the optimal rate (cf.~\eqref{eq:optimal-first-order-rate}) for any channels that are not necessarily full rank.
In fact, we derive even stronger result, i.e., the strong converse theorem.

Once we have established the strong converse theorem, the next step is the second-order
asymptotics \cite{Strassen:62,hayashi:09,polyanskiy:10}. In this paper, we also derive the optimal 
second-order rate of channel resolvability under a condition (cf.~\eqref{eq:condition}). 
Furthermore, under the condition that
a DMC has the unique capacity achieving input distribution, 
we derive the optimal second-order rate of 
channel resolvability for the worst input distribution.

The rest of this paper is organized as follows: we introduce the problem setting of channel resolvability and main results 
in Section \ref{section:formulation-results}. Then, we will show proofs of main results in Section \ref{section:proofs}.
We conclude in Section \ref{section:conclusion} and discuss open problems.
The proofs of technical lemmas are given in appendices.

\section{Formulation and Results} \label{section:formulation-results}

\subsection{Problem Formulation}

For a given input distribution $p_n \in {\cal P}({\cal X}^n)$ on ${\cal X}^n$ and 
a given channel $W: x \mapsto W_x$, the goal of the channel resolvability problem 
(for DMCs) is
to approximate the output distribution
\begin{eqnarray*}
W_{p_n}(\bm{y}) := \sum_{\bm{x} \in {\cal X}^n} p_n(\bm{x}) W^n_{\bm{x}}(\bm{y}),
\end{eqnarray*}
$W_{\bm{x}}^n(\bm{y}) = W_{x_1}(y_1)\cdots W_{x_n}(y_n)$ is the $n$th independent extension of $W$
with input vector $\bm{x}$.
Throughout the paper, we assume that alphabets are finite.
More precisely, a channel resolvability code ${\cal C}_n$ of size $|{\cal C}_n| = M_n$ is 
a set of codewords ${\cal C}_n = \{\bm{x}_1,\ldots,\bm{x}_{M_n} \} \subset {\cal X}^n$,
and we are interested in approximating $W_{p_n}$ by
\begin{eqnarray*}
W_{{\cal C}_n} := \sum_{i=1}^{M_n} \frac{1}{M_n} W_{\bm{x}_i}^n.
\end{eqnarray*}
In this paper, the approximation error is evaluated by the normalized 
variational distance:
\begin{eqnarray*}
\rho({\cal C}_n,W_{p_n}) := \frac{1}{2} \| W_{{\cal C}_n} - W_{p_n} \|_1.
\end{eqnarray*}
For a given $0 \le \varepsilon < 1$, we define the minimum size of the random number needed to approximate $W_{p_n}$ by 
\begin{eqnarray*}
R(n,\varepsilon | p_n) := \inf_{{\cal C}_n}\left\{ \frac{1}{n} \log |{\cal C}_n| : \rho({\cal C}_n,W_{p_n}) \le \varepsilon \right\}.
\end{eqnarray*}
We also consider the worst input distribution case:
\begin{eqnarray*}
R_{\san{wst}}(n,\varepsilon) := \sup\{ R(n,\varepsilon |p_n): p_n \in {\cal P}({\cal X}^n) \},
\end{eqnarray*}
where the supremum is taken over all distributions on ${\cal X}^n$ that are not
necessarily i.i.d.

\subsection{Fixed I.I.D. Input Distribution}

First, we consider the case in which the input distribution is fixed as 
$p_n = p^n$ for $n$th i.i.d. extension of $p \in {\cal P}({\cal X})$.
When the transition vectors $\{ W_x \}_{x \in {\cal X}}$ are linearly independent, 
the channel $W$ is called {\em full rank}. For full rank channels, the following result is known.
\begin{proposition}[\cite{han:93, han:93b}] \label{proposition:weak-converse}
For a full rank channel\footnote{The full rank condition is only needed in the converse part \cite{han:93b}.}, we have
\begin{eqnarray}
\lim_{\varepsilon \downarrow 0} \limsup_{n\to\infty} R(n,\varepsilon |p^n) = I(p,W),
\end{eqnarray}
where $I(p,W)$ is the mutual information for the input distribution $p$.
\end{proposition}

When a channel is not necessarily full rank, more than one $q \in {\cal P}({\cal X})$
satisfying $W_q = W_p$ may exist. Thus, we introduce the following quantity:
\begin{eqnarray} \label{eq:optimal-first-order-rate}
S_{W_p} := \min\left\{ I(q,W) : q \in {\cal P}({\cal X}), W_q = W_p \right\}.
\end{eqnarray}
In general, $S_{W_p}$ is strictly smaller than $I(p,W)$, as is illustrated by the following example.
\begin{example}[\cite{han:93}] \label{example:non-full-rank}
For ${\cal X} = \{0,1,\san{e}\}$ and ${\cal Y} = \{0,1\}$, let $W$ be given by
\begin{eqnarray*}
W_0(0) = 1,~~W_1(1)=1,~~W_{\san{e}}(0) = W_{\san{e}}(1) = 1/2.
\end{eqnarray*}
Let $p$ be such that $p(0) = p(1) = 1/2$. Then, we have $I(p,W) = 1$
but $S_{W_p} = 0$.
\end{example}

We can derive the following refinement of Proposition \ref{proposition:weak-converse}.
\begin{theorem}[First Order Asymptotics for Fixed $p$] \label{theorem:first-order}
For any $0 < \varepsilon < 1$, we have
\begin{eqnarray}
\lim_{n\to\infty} R(n,\varepsilon | p^n) = S_{W_p}.
\end{eqnarray}
\end{theorem}

Fon an input distributions $q$, let
\begin{eqnarray*}
U_{q,W} := \sum_{x,y} q(x) W_x(y)\left[ \log \frac{W_x(y)}{W_q(y)} - I(q,W) \right]^2
\end{eqnarray*}
and
\begin{eqnarray*}
V_{q,W_p} := \sum_{x,y} q(x) W_x(y) \left[  \log \frac{W_x(y)}{W_p(y)} - D(W_x \| W_p) \right]^2,
\end{eqnarray*}
where $D(\cdot \| \cdot)$ is the KL divergence. For $q$ satisfying $W_q = W_p$, 
$U_{q,W}$ and $V_{q,W_p} = V_{q,W_q}$ are the unconditional information variance and conditional information variance
respectively \cite{polyanskiy:10}. In such a case, we have
\begin{eqnarray}
V_{q,W_p} \le U_{q,W},
\end{eqnarray}
and the equality hold if and only if
\begin{eqnarray*}
D(W_x \| W_p) = I(q,W)~\forall x \mbox{ s.t. } q(x) > 0.
\end{eqnarray*}
Let 
\begin{eqnarray*}
{\cal V}(p,W) := \left\{ q \in {\cal P}({\cal X}) : I(q,W) = S_{W_p}, W_q = W_p \right\}.
\end{eqnarray*}
Then, we define the following four quantities: 
\begin{eqnarray}
U_{p,W}^+ &:=& \max_{q \in {\cal V}(p,W)} U_{q,W}, \\
U_{p,W}^- &:=& \min_{q \in {\cal V}(p,W)} U_{q,W}, \\
V_{p,W}^+ &:=& \max_{q \in {\cal V}(p,W)} V_{q,W_p}, \label{eq:upper-conditional-variance} \\
V_{p,W}^- &:=& \min_{q \in {\cal V}(p,W)} V_{q,W_p}. \label{eq:lower-conditional-variance}
\end{eqnarray}
\begin{theorem}[Second Order Asymptotics for Fixed $p$] \label{theorem:second-order}
We have
\begin{eqnarray}
\lefteqn{ \limsup_{n \to \infty} \sqrt{n} \left(R(n,\varepsilon |p^n) - S_{W_p} \right) } \nonumber \\
 &\le& \left\{
 \begin{array}{ll}
 \sqrt{U_{p,W}^+} Q^{-1}(\varepsilon) & \varepsilon \ge 1/2 \\
 \sqrt{U_{p,W}^-} Q^{-1}(\varepsilon) & \varepsilon < 1/2
 \end{array}
 \right.
 \label{eq:second-direct}
\end{eqnarray}
provided that $U_{p,W}^- > 0$.
Furthermore, if 
\begin{eqnarray} \label{eq:condition}
D(W_x \| W_p) = S_{W_p}~\forall x \in {\cal X}
\end{eqnarray}
and $V_{p,W}^- > 0$ hold, we have
\begin{eqnarray}
\lefteqn{ \lim_{n\to\infty} \sqrt{n} \left(R(n,\varepsilon) - S_{W_p} \right) } \nonumber \\
 &=& \left\{
 \begin{array}{ll}
 \sqrt{V_{p,W}^+} Q^{-1}(\varepsilon) & \varepsilon \ge 1/2 \\
 \sqrt{V_{p,W}^-} Q^{-1}(\varepsilon) & \varepsilon < 1/2
 \end{array}
 \right.
 \label{eq:second-converse} \\
 &=& \left\{
 \begin{array}{ll}
 \sqrt{U_{p,W}^+} Q^{-1}(\varepsilon) & \varepsilon \ge 1/2 \\
 \sqrt{U_{p,W}^-} Q^{-1}(\varepsilon) & \varepsilon < 1/2
 \end{array}
 \right.,
 \label{eq:second-converse-2}
\end{eqnarray}
where 
\begin{eqnarray*}
Q(a) := \int_{a}^\infty \frac{1}{\sqrt{2\pi}} \exp\left[- \frac{t^2}{2}\right] dt.
\end{eqnarray*}
\end{theorem}


\begin{remark} \label{remark-1}
In the converse part, we are going to prove the inequality $\ge$ in \eqref{eq:second-converse}.
It should be noted that the condition in \eqref{eq:condition} is not only used as a matching 
condition for \eqref{eq:second-converse} and \eqref{eq:second-converse-2}
to coincide, but it is crucially used in the converse proof.
In fact, the inequality $\ge$ in \eqref{eq:second-converse} does not hold in general since
the inequality
\begin{eqnarray*}
\sqrt{V_{p,W}^+} Q^{-1}(\varepsilon) > \sqrt{U_{p,W}^+} Q^{-1}(\varepsilon)
\end{eqnarray*}
may hold for $\varepsilon > 1/2$, which contradicts the achievability part.
\end{remark}

\begin{remark} \label{remark-2}
When channel $W$ is a noiseless channel, the channel resolvability problem reduces to the source resolvability 
problem \cite[Sec.~2]{han:book}. In this case, since the channel is full rank, ${\cal V}(p,W)$ is the singleton $\{p\}$. 
We also have $S_{W_p} = H( p)$, $V_{p,W}^+ = V_{p,W}^- = 0$,  and 
$U_{p,W}^* := U_{p,W}^+ = U_{p,W}^-$. Although this case is not covered by Theorem \ref{theorem:second-order},
the second order asymptotics for this case is already known to be \cite{nomura:13}
\begin{eqnarray*}
\limsup_{n \to \infty} \sqrt{n} \left(R(n,\varepsilon |p^n) - H(p ) \right) = \sqrt{U_{p,W}^*} Q^{-1}(\varepsilon).
\end{eqnarray*}
\end{remark}


\subsection{Worst Input Distribution}

Next, we consider the worst input distribution case. Let 
\begin{eqnarray*}
C_W := \max\{ I(p,W) : p \in {\cal P}({\cal X}) \}
\end{eqnarray*}
be the channel capacity of $W$. 
The following result is known.
\begin{proposition}[\cite{han:93}]
For any $0 < \varepsilon < 1$, we have
\begin{eqnarray*}
\lim_{n\to\infty} R_{\san{wst}}(n,\varepsilon) = C_W.
\end{eqnarray*}
\end{proposition}

Let 
\begin{eqnarray*}
{\cal V}(W) := \{ p \in {\cal P}({\cal X}) : I(p,W) = C_W \}
\end{eqnarray*}
be the set of all capacity achieving input distribution (CAID).
It is well known that the output distribution $W_{p^*}$ for any CAID $p^*$ is unique. 
Let us introduce {\em full support CAID condition}:
\begin{eqnarray} \label{eq:full-support-caid}
D(W_x \| W_{p^*}) = C_W~~~\forall x \in {\cal X}.
\end{eqnarray} 
Under this condition, we find that 
\begin{eqnarray} \label{eq:equality-iid-resolvability-capacity}
S_{W_{p^*}} = C_W
\end{eqnarray}
holds. Moreover, $V_{p^*,W}^+$ and $V_{p^*,W}^-$ defined in \eqref{eq:upper-conditional-variance}
and \eqref{eq:lower-conditional-variance}
coincide with the conditional variances that appear in the channel coding problems:
\begin{eqnarray*}
V_W^+ &:=& \max_{p \in {\cal V}(W)} V_{p,W_p}, \\
V_W^- &:=& \min_{p \in {\cal V}(W)} V_{p,W_p}.
\end{eqnarray*}

\begin{theorem}[Second Order Asymptotics for the Worst Case] \label{theorem-worst-case}
Suppose that the full support CAID condition is satisfied (cf.~\eqref{eq:full-support-caid}). Then, we have
\begin{eqnarray}
\lefteqn{
\limsup_{n\to\infty} \sqrt{n}\left(R_{\san{wst}}(n,\varepsilon) - C_W \right)
} \nonumber \\
&\le& 
\left\{
\begin{array}{ll}
\sqrt{V_W^-} Q^{-1}(\varepsilon) & \varepsilon \ge 1/2 \\
\sqrt{V_W^+} Q^{-1}(\varepsilon) & \varepsilon < 1/2
\end{array}
\right.
\end{eqnarray}
and
\begin{eqnarray}
\lefteqn{
\liminf_{n\to\infty} \sqrt{n}\left(R_{\san{wst}}(n,\varepsilon) - C_W \right)
} \nonumber \\
&\ge& 
\left\{
\begin{array}{ll}
\sqrt{V_W^+} Q^{-1}(\varepsilon) & \varepsilon \ge 1/2 \\
\sqrt{V_W^-} Q^{-1}(\varepsilon) & \varepsilon < 1/2
\end{array}
\right.
\end{eqnarray}
provided that $V_W^- > 0$.
\end{theorem}

\begin{remark} \label{remark:full-support-caid}
It should be noted that \eqref{eq:equality-iid-resolvability-capacity} is not true in general.
In fact, the channel in Example \ref{example:non-full-rank} does not satisfy \eqref{eq:equality-iid-resolvability-capacity}.
It should be also noted that 
\eqref{eq:equality-iid-resolvability-capacity} is slightly weaker condition than \eqref{eq:full-support-caid}.
These conditions are needed only in the converse part, and 
for the achievability part of Theorem \ref{theorem-worst-case}, we need not to assume neither
\eqref{eq:full-support-caid} nor \eqref{eq:equality-iid-resolvability-capacity}.
\end{remark}

\section{Proofs of Main Results} \label{section:proofs}

\subsection{Preliminaries for Proofs}

The purpose of this section is to prepare lemmas that will be used for the achievability part and the converse part, respectively.
To save space, we introduce a notation that is usually used in quantum information (eg.~\cite{nagaoka:07}).
For a function $A$ on ${\cal Y}$, let $\{ A \ge 0 \}$ indicates the set $\{ y : A(y) \ge 0 \}$.
Then, for a non-negative function $P$ on ${\cal Y}$ (not necessarily normalized), we denote
$P\{A\ge0 \} := \sum_{y \in \{A\ge0\}} P(y)$.

The following lemma guarantees existence of a good channel resolvability code.
\begin{lemma}[Theorem 2 of \cite{hayashi:06b}] \label{lemma:achievability}
For any $q_n \in {\cal P}({\cal X}^n)$ such that $W_{q_n} = W_{p_n}$ 
and any real number $C_n$, there exists a channel resolvability code ${\cal C}_n$ such that
\begin{eqnarray*}
\lefteqn{ \rho({\cal C}_n,W_{p_n}) } \\
&\le& \sum_{\bm{x}} q_n(\bm{x}) W_{\bm{x}}^n\left\{ W_{\bm{x}}^n - C_n W_{p_n} \ge 0 \right\} + \frac{1}{2} \sqrt{\frac{C_n}{M_n}}.
\end{eqnarray*}
\end{lemma}

In the converse part, we are going to use the argument of the typical sequence.
Let $T_{p,\delta}$ be the set of typical sequences, i.e.,
$|P_{\bm{x}}(a) - p(a) | \le \delta~\forall a \in {\cal X}$
and, in addition, no $a \in {\cal X}$ with $p(a) = 0$ occur in $\bm{x}$, 
where $P_{\bm{x}}$ is the type of sequence $\bm{x}$. 
We also define the set $T_{W,\delta}(\bm{x})$ of $W$-typical sequences given $\bm{x}$, i.e., 
$| P_{\bm{x}\bm{y}}(a,b) - P_{\bm{x}}(a) W_a(b)| \le \delta~\forall (a,b) \in {\cal X} \times {\cal Y}$
and, in addition, $P_{\bm{x}\bm{y}}(a,b) = 0$ whenever $W_a(b) = 0$,
where $P_{\bm{x}\bm{y}}$ is the joint type of $(\bm{x},\bm{y})$.
For the output distribution, we also define the set of typical sequences: $T_{W_p,\delta}$.
For any $\delta> 0$, it is well known that \cite[Lemma 2.12]{csiszar-korner:11}
\begin{eqnarray*}
p^n(T_{p,\delta}) &\ge& 1 - \gamma_n, \\
W_p^n(T_{W_p,\delta}) &\ge& 1 - \gamma_n, \\
W_{\bm{x}}^n(T_{W,\delta}(\bm{x})) &\ge& 1 - \gamma_n~\forall \bm{x} \in {\cal X}^n
\end{eqnarray*}
for some $\gamma_n$ such that $\gamma_n \to 0$ as $n\to \infty$.

Let
\begin{eqnarray*}
{\cal A}_n(\delta) := \left\{ \bm{x} : |W_{P_{\bm{x}}}(b) - W_p(b) | > 2|{\cal X}| \delta \mbox{ for some } b \in {\cal Y} \right\}
\end{eqnarray*}
be the set of all sequences such that the output distribution $W_{P_{\bm{x}}}$ is not close to $W_p$.
For such sequences, we have the following property.
\begin{lemma} \label{lemma:output-of-non-typical}
For $\bm{x} \in {\cal A}_n(\delta)$, we have $T_{W,\delta}(\bm{x}) \subset T_{W_p,\delta^\prime}^c$ for $\delta^\prime = |{\cal X}|\delta$.
\end{lemma} 

The following will be used as a key lemma in the converse part. 
\begin{lemma} \label{lemma:converse-key}
For a given channel resolvability code ${\cal C}_n$, let ${\cal B}_n = \{ i : \bm{x}_i \in {\cal A}_n(\delta) \}$. 
Then, for any $\alpha \ge 0$ and sufficiently large $n$, we have
\begin{eqnarray*}
\lefteqn{ \rho({\cal C}_n, W_p^n) } \\
&\ge& \frac{|{\cal B}_n|}{M_n}(1-\gamma_n) + \sum_{i \in {\cal B}_n^c} \frac{1}{M_n} W_{\bm{x}_i}^n\{ W_{\bm{x}_i}^n - e^\alpha M_n W_p^n \ge 0\}  \\
&&  - e^{-\alpha} - \gamma_n 
\end{eqnarray*}
for some $\gamma_n$ such that $\gamma_n \to 0$ as $n\to \infty$.
\end{lemma}

The following two lemmas are also used in the converse part.
\begin{lemma} \label{lemma:perturbation-expectation}
Suppose $\bm{x} \notin {\cal A}_n(\delta)$. Then, we have
\begin{eqnarray*}
\sum_a P_{\bm{x}}(a) D(W_a \| W_p) + \tau(\delta) \ge S_{W_p}
\end{eqnarray*}
for some $\tau(\delta)$ such that $\tau(\delta) \to 0$ as $\delta \to 0$.
\end{lemma}

\begin{lemma} \label{lemma:perturbation-variance}
Suppose \eqref{eq:condition} holds and $\bm{x} \notin {\cal A}_n(\delta)$. Then, we have
\begin{eqnarray}
V_{P_{\bm{x}},W_p} + \tau_1(\delta) &\ge& V_{p,W}^-, \label{eq:perturbation-variance-1} \\
V_{P_{\bm{x}},W_p} - \tau_2(\delta) &\le& V_{p,W}^+ \label{eq:perturbation-variance-2}
\end{eqnarray}
for some $\tau_1(\delta)$ and $\tau_2(\delta)$ that converge to $0$ as $\delta \to 0$.
\end{lemma}

\subsection{Proofs of Theorem \ref{theorem:first-order}}

\paragraph*{Direct Part}
Let $q$ be such that $I(q,W) = S_{W_p}$. For arbitrarily fixed $\nu > 0$, we use Lemma \ref{lemma:achievability} by
setting $M_n = e^{n(I(q,W)+2\nu)}$ and $C_n = e^{n (I(q,W)+\nu)}$. Then, by the law of large number,
we have $\rho({\cal C}_n, W_p^n) \to 0$. Since $\nu >0$ can be arbitrary, we complete the proof. \qed

\paragraph*{Converse Part}

For arbitrary $0 < \varepsilon < 1$, suppose 
\begin{eqnarray*}
\liminf_{n \to \infty} R(n,\varepsilon |p^n) < S_{W_p}.
\end{eqnarray*}
Then, there exist $\nu > 0$ and a code ${\cal C}_n$ such that $\rho({\cal C}_n)\le\varepsilon$ and
\begin{eqnarray} \label{eq:assumption-of-contradiction}
\frac{1}{n} \log M_n \le S_{W_p} - 3 \nu
\end{eqnarray}
for infinitely many $n$. For $q \in {\cal P}({\cal X})$, we denote 
\begin{eqnarray*}
D(W \| W_p|q) := \sum_a q(a) D(W_a \| W_p).
\end{eqnarray*}
From Lemma \ref{lemma:perturbation-expectation}, if we take $\delta$ sufficiently small, we have
\begin{eqnarray} \label{eq:proof-first-order-perturbation}
D(W\|W_p|P_{\bm{x}}) \ge S_{W_p} - \nu
\end{eqnarray} 
for every $\bm{x} \notin {\cal A}_n(\delta)$. 

By applying Lemma \ref{lemma:converse-key} for $\alpha = \nu n$, we have
\begin{eqnarray}
\lefteqn{ \rho({\cal C}_n, W_p^n)}  \nonumber \\
&\ge& \frac{|{\cal B}_n|}{M_n}(1-\gamma_n) + \sum_{i \in {\cal B}_n^c} \frac{1}{M_n} W_{\bm{x}_i}^n\{ W_{\bm{x}_i}^n - e^{\nu n} M_n W_p^n \ge 0\}  \nonumber \\
&&  - e^{- \nu n} - \gamma_n.
\label{eq:first-order-converse-1}
\end{eqnarray}
Here, the third term and the forth term converge to $0$. From \eqref{eq:assumption-of-contradiction}, the second term is further lower bounded by
\begin{eqnarray*}
\lefteqn{ \sum_{i \in {\cal B}_n^c} \frac{1}{M_n} W_{\bm{x}_i}^n\left\{ \frac{1}{n} \log \frac{W_{\bm{x}_i}^n}{W_p^n}  \ge S_{W_p} - 2 \nu \right\} } \\
&\stackrel{(\rom{a})}{\ge}& \sum_{i \in {\cal B}_n^c} \frac{1}{M_n} W_{\bm{x}_i}^n\left\{ \frac{1}{n} \log \frac{W_{\bm{x}_i}^n}{W_p^n}  \ge D(W\|W_p|P_{\bm{x}_i}) -  \nu \right\}, 
\end{eqnarray*}
where $(\rom{a})$ follows from \eqref{eq:proof-first-order-perturbation}.
Here, note that
\begin{eqnarray} \label{eq:converse-expectation}
\mathsf{E}_{W_{\bm{x}_i}^n}\left[ \frac{1}{n} \log \frac{W_{\bm{x}_i}^n(\bm{Y})}{W_p^n(\bm{Y})} \right] = D(W\|W_p|P_{\bm{x}_i})
\end{eqnarray}
and
\begin{eqnarray} \label{eq:converse-variance}
\mathsf{V}_{W_{\bm{x}_i}^n} \left[ \frac{1}{n} \log \frac{W_{\bm{x}_i}^n(\bm{Y})}{W_p^n(\bm{Y})} \right] &=& \frac{V_{P_{\bm{x}_i},W_p}}{n} \\
&\le& \frac{\max_q V_{q,W_p}}{n},
\end{eqnarray}
where $\mathsf{E}_{W_{\bm{x}_i}^n}$ and $\mathsf{V}_{W_{\bm{x}_i}^n}$ are the expectation and the variance with
respect to $\bm{Y} \sim W_{\bm{x}_i}^n$. Thus, by using Chebyshev's inequality, we have
\begin{eqnarray*}
&& W_{\bm{x}_i}^n\left\{ \frac{1}{n} \log \frac{W_{\bm{x}_i}^n}{W_p^n}  \ge D(W\|W_p|P_{\bm{x}_i}) -  \nu \right\}  \\
&&~~~\ge 1 - \frac{\max_q V_{q,W_p}}{\nu^2 n}.
\end{eqnarray*}
Consequently, from \eqref{eq:first-order-converse-1},
we have $\rho({\cal C}_n, W_p^n) \to 1$, which contradict with $\rho({\cal C}_n, W_p^n) \le \varepsilon$. Thus, we have
$\liminf_{n \to \infty} R(n,\varepsilon) \ge S_{W_p}$.
\qed

\subsection{Proofs of Theorem \ref{theorem:second-order}}

\paragraph*{Direct Part}
Let $q$ be such that $I(q,W) = S_{W_p}$ and $U_{q,W} = U_{q,W}^-$ (or $U_{q,W} = U_{q,W}^+$).
For arbitrarily fixed $\nu > 0$, we use Lemma \ref{lemma:achievability} by setting
$\log M_n = n I(q,W) + \sqrt{n U_{q,W}} Q^{-1}(\varepsilon-\nu) + \log n$ 
and $\log C_n = n I(q,W) + \sqrt{n U_{q,W}} Q^{-1}(\varepsilon-\nu)$. Then, by the central limit theorem, we have
$\rho({\cal C}_n, W_p^n) \le \varepsilon$ for sufficiently large $n$. Since $\nu > 0$ can be arbitrary, we complete the proof of \eqref{eq:second-direct}. \qed

\paragraph*{Converse Part}

We only prove\footnote{For $\varepsilon > 1/2$, we replace 
$V_{p,W}^-$ in \eqref{eq:second-order-converse-proof-2} by $V_{p,W}^+$, which follows from \eqref{eq:perturbation-variance-2} of
Lemma \ref{lemma:perturbation-variance} by noting $Q^{-1}(\varepsilon) < 0$ for $\varepsilon > 1/2$.} the case with $\varepsilon < 1/2$. Suppose 
\begin{eqnarray*}
\liminf_{n\to\infty} \sqrt{n}\left( R(n,\varepsilon |p^n) - nS_{W_p} \right) < \sqrt{V_{p,W}^-} Q^{-1}(\varepsilon).
\end{eqnarray*}
Then, there exists $\nu > 0$ and a code ${\cal C}_n$ such that $\rho({\cal C}_n) \le \varepsilon$ and
\begin{eqnarray} \label{eq:second-order-converse-proof-1}
\log M_n \le n S_{W_p} + \sqrt{n V_{p,W}^-} Q^{-1}(\varepsilon) - 3 \nu \sqrt{n}
\end{eqnarray}
for infinitely many $n$. From \eqref{eq:perturbation-variance-1} of Lemma \ref{lemma:perturbation-variance}, if we take $\delta$ sufficiently small, we have
\begin{eqnarray} \label{eq:second-order-converse-proof-2}
\sqrt{V_{P_{\bm{x}},W_p} } Q^{-1}(\varepsilon) \ge \sqrt{ V_{p,W}^- } Q^{-1}(\varepsilon) - \nu
\end{eqnarray}
for ever $\bm{x} \notin {\cal A}_n(\delta)$.

By applying Lemma \ref{lemma:converse-key} for $\alpha = \nu \sqrt{n}$, we have
\begin{eqnarray}
\lefteqn{ \rho({\cal C}_n, W_p^n) \ge }  \nonumber \\
&& \hspace{-1mm} \frac{|{\cal B}_n|}{M_n}(1-\gamma_n) + \sum_{i \in {\cal B}_n^c} \frac{1}{M_n} W_{\bm{x}_i}^n\{ W_{\bm{x}_i}^n - e^{\nu \sqrt{n}} M_n W_p^n \ge 0\}  \nonumber \\
&&  - e^{- \nu \sqrt{n}} - \gamma_n.
\label{eq:second-order-converse-proof-3}
\end{eqnarray}
From \eqref{eq:second-order-converse-proof-1}, each term in the summation of the second term is further lower bounded by
\begin{eqnarray}
\lefteqn{ W_{\bm{x}_i}^n\left\{ \frac{1}{\sqrt{n}}\left( \log \frac{W_{\bm{x}_i}^n}{W_p^n} -nS_{W_p}\right) \ge \sqrt{ V_{p,W}^-} Q^{-1}(\varepsilon) - 2\nu  \right\}  } 
 \nonumber \\
&\stackrel{(\rom{a})}{\ge}& \nonumber \\
&& \hspace{-11mm} W_{\bm{x}_i}^n\left\{ \frac{1}{\sqrt{n}}\left( \log \frac{W_{\bm{x}_i}^n}{W_p^n} -nS_{W_p}\right) \ge \sqrt{V_{P_{\bm{x}_i},W_p}} Q^{-1}(\varepsilon) - \nu  \right\}, \nonumber \\
\label{eq:second-order-converse-proof-4}
\end{eqnarray}
where $(\rom{a})$ follows from \eqref{eq:second-order-converse-proof-2}.
Here, we note that $D(W\|W_p|P_{\bm{x}}) = S_{W_p}$
holds for any sequence $\bm{x}$ because of the assumption in \eqref{eq:condition}.
Now, by noting \eqref{eq:converse-expectation} and \eqref{eq:converse-variance}, and by using the central limit theorem,
\eqref{eq:second-order-converse-proof-4} is strictly larger than $\varepsilon$ for sufficiently large $n$. Thus, from \eqref{eq:second-order-converse-proof-3},
we have $\rho({\cal C}_n, W_p^n) > \varepsilon$ for sufficiently large $n$, which is a contradiction. 
Thus, we have
\begin{eqnarray*}
\liminf_{n\to\infty} \sqrt{n}\left( R(n,\varepsilon |p^n) - nS_{W_p} \right) \ge \sqrt{V_{p,W}^-} Q^{-1}(\varepsilon),
\end{eqnarray*}
which completes the proof of $\ge$ in \eqref{eq:second-converse}.
The equality between \eqref{eq:second-converse} and \eqref{eq:second-converse-2} follows from the assumption in \eqref{eq:condition}.
\qed


\subsection{Proof of Theorem \ref{theorem-worst-case}}

\paragraph*{Direct Part}
Let $p^*$ be CAID, and let $V_W = V_W^+$ when $\varepsilon < 1/2$ (or $V_W^-$ when $\varepsilon \ge 1/2$).
From Lemma \ref{lemma:achievability} with $q_n = p_n$, there exists a resolvability code satisfying
\begin{eqnarray*}
\lefteqn{
\rho({\cal C}_n,W_{p_n})
} \\
&\le& \sum_{\bm{x}} p_n(\bm{x}) W_{\bm{x}}^n\left\{ \log \frac{W_{\bm{x}}^n}{W_{p_n}} \ge \log C_n \right\} 
+ \frac{1}{2} \sqrt{\frac{C_n}{M_n}}.
\end{eqnarray*}
Here, by the change of measure argument, we have
\begin{eqnarray*}
\lefteqn{ 
W_{\bm{x}}^n\left\{ \log \frac{W_{\bm{x}}^n}{W_{p_n}} \ge \log C_n \right\} 
} \\
&=& W_{\bm{x}}^n\left\{ \log \frac{W_{\bm{x}}^n}{W_{p^*}^n} + \log \frac{W_{p^*}^n}{W_{p_n}} \ge \log C_n \right\} \\
&\le& W_{\bm{x}}^n\left\{ \log \frac{W_{\bm{x}}^n}{W_{p^*}^n} \ge \log C_n - \xi \right\}
 + W_{\bm{x}}^n\left\{ \log \frac{W_{p^*}^n}{W_{p_n}} \ge \xi \right\}
\end{eqnarray*}
for any $\xi > 0$, which implies 
\begin{eqnarray}
\lefteqn{
\sum_{\bm{x}} p_n(\bm{x}) W_{\bm{x}}^n\left\{ \log \frac{W_{\bm{x}}^n}{W_{p_n}} \ge \log C_n \right\}
} \nonumber \\
&\le& \hspace{-3mm}
\sum_{\bm{x}} p_n(\bm{x}) W_{\bm{x}}^n\left\{ \log \frac{W_{\bm{x}}^n}{W_{p^*}^n} \ge \log C_n - \xi \right\} + e^{-\xi}.
\label{eq:proof-worst-case-bound-1}
\end{eqnarray}
Now, for arbitrarily fixed $\nu > 0$, let $\xi = \log n$, $\log M_n = n C_W + \sqrt{n V_W} Q^{-1}(\varepsilon - \nu) + 2 \log n$
and $\log C_n = n C_W + \sqrt{n V_W} Q^{-1}(\varepsilon -\nu) + \log n$. Then,
by applying the central limit theorem for each $W_{\bm{x}}^n \{ \cdot \}$ in \eqref{eq:proof-worst-case-bound-1},
we have $\rho({\cal C}_n,W_{p_n}) \le \varepsilon$ for sufficiently large $n$.
Since $\nu > 0$ can be arbitrary, we complete the proof of the direct part. \qed

\paragraph*{Converse Part}
From the definition of the worst case, we have
\begin{eqnarray*}
R_{\san{wst}}(n,\varepsilon) \ge R(n,\varepsilon| (p^*)^n).
\end{eqnarray*}
Thus, the converse part follows from Theorem \ref{theorem:second-order}.

\section{Conclusion} \label{section:conclusion}

As we discussed in Remark \ref{remark-1}, the optimal second-order rate 
for fixed i.i.d. input distribution is not clear in general.
One possible answer is that the optimal second-order rate is always given by \eqref{eq:second-converse-2}.
This is at least true for noiseless channel (cf.~Remark \ref{remark-2}), but there is no strong evidence in general.
Clarifying the optimal second-order rate is an important future research agenda.
There is also a gap between the achievability and the converse for the worst input distribution case
in general (cf.~Theorem \ref{theorem-worst-case}); the gap vanishes only when the channel has 
the unique CAID. 


\appendix

\subsection{Proof of Lemma \ref{lemma:output-of-non-typical}}

From the definition of $T_{W,\delta}(\bm{x})$, $\bm{y} \in T_{W,\delta}(\bm{x})$ implies 
$ |P_{\bm{y}}(b) - W_{P_{\bm{x}}}(b) | \le \delta^\prime~\forall b \in {\cal Y}$.
On the other hand, from the definition of ${\cal A}_n(\delta)$, there exists $b \in {\cal Y}$ such that
\begin{eqnarray} \label{eq:y-deviate}
|W_{P_{\bm{x}}}(b) - W_p(b) | > 2 \delta^\prime.
\end{eqnarray}
Thus, for $b$ satisfying \eqref{eq:y-deviate}, $\bm{y} \in T_{W,\delta}(\bm{x})$ implies
\begin{eqnarray*}
\lefteqn{ | P_{\bm{y}}(b) - W_p(b) | } \\
&\ge& |W_{P_{\bm{x}}}(b) - W_p(b)| - |P_{\bm{y}}(b) - W_{P_{\bm{x}}}(b) | \\
&>& \delta^\prime,
\end{eqnarray*}
which implies $\bm{y} \notin T_{W_p,\delta^\prime}$.
\qed

\subsection{Proof of Lemma \ref{lemma:converse-key}}

First, we divide $W_p^n$ into typical part and non-typical part as
$W_p^n = \hat{W}_p^n + \tilde{W}_p^n$, where
\begin{eqnarray*}
\hat{W}_p^n(\bm{y}) &:=& W_p^n(\bm{y}) \mathbf{1}[\bm{y} \in T_{W_p,\delta^\prime}], \\
\tilde{W}_p^n(\bm{y}) &:=& W_p^n(\bm{y}) \mathbf{1}[\bm{y} \notin T_{W_p,\delta^\prime}],
\end{eqnarray*}
where $\delta^\prime$ is specified in Lemma \ref{lemma:output-of-non-typical}, and $\mathbf{1}[\cdot]$ is the indicator function.
Then,  for sufficiently large $n$, we have
\begin{eqnarray}
\lefteqn{ \frac{1}{2} \| W_{{\cal C}_n} - W_p^n \|_1 } \nonumber \\
&\stackrel{(\rom{a})}{\ge}& W_{{\cal C}_n} \{ W_{{\cal C}_n} - e^\alpha \hat{W}_p^n \ge 0 \} - W_p^n\{ W_{{\cal C}_n} - e^\alpha \hat{W}_p^n \ge 0 \} \nonumber \\
&\ge&  W_{{\cal C}_n} \{ W_{{\cal C}_n} - e^\alpha \hat{W}_p^n \ge 0 \} - \hat{W}_p^n\{ W_{{\cal C}_n} - e^\alpha \hat{W}_p^n \ge 0 \} \nonumber \\
 && - \tilde{W}_p^n({\cal Y}^n) \nonumber \\
&\stackrel{(\rom{b})}{\ge}& W_{{\cal C}_n} \{ W_{{\cal C}_n} - e^\alpha \hat{W}_p^n \ge 0 \} - e^{-\alpha} - \gamma_n,
\label{eq:proof-key-lemma-0}
\end{eqnarray}
where $(\rom{a})$ follows form the definition of the variational distance, and $(\rom{b})$ follows from 
\begin{eqnarray*}
\hat{W}_p^n\{ W_{{\cal C}_n} - e^\alpha \hat{W}_p^n \ge 0 \}
 &\le& e^{-\alpha} W_{{\cal C}_n} \{ W_{{\cal C}_n} - e^\alpha \hat{W}_p^n \ge 0 \} \\
 &\le& e^{-\alpha}
\end{eqnarray*}
and $\tilde{W}_p^n({\cal Y}^n) = W_p^n(T_{W_p,\delta^\prime}^c) \le \gamma_n$ for sufficiently large $n$.

Furthermore, we have
\begin{eqnarray}
\lefteqn{ W_{{\cal C}_n} \{ W_{{\cal C}_n} - e^\alpha \hat{W}_p^n \ge 0 \} } \nonumber \\
&=& \sum_{i=1}^{M_n} \frac{1}{M_n} W_{\bm{x}_i}^n \left\{ \sum_{j=1}^{M_n} W_{\bm{x}_j}^n - e^\alpha M_n \hat{W}_p^n \ge 0 \right\} \nonumber \\
&\stackrel{(\rom{c})}{\ge}& \sum_{i=1}^{M_n} \frac{1}{M_n} W_{\bm{x}_i}^n \left\{  W_{\bm{x}_i}^n - e^\alpha M_n \hat{W}_p^n \ge 0\right\} \nonumber \\
&=& \sum_{i \in {\cal B}_n} \frac{1}{M_n} W_{\bm{x}_i}^n \left\{  W_{\bm{x}_i}^n - e^\alpha M_n \hat{W}_p^n \ge 0\right\} \nonumber \\
&& + \sum_{i \in {\cal B}_n^c} \frac{1}{M_n} W_{\bm{x}_i}^n \left\{  W_{\bm{x}_i}^n - e^\alpha M_n \hat{W}_p^n \ge 0\right\}, 
 \label{eq:proof-key-lemma-1}
\end{eqnarray}
where $(\rom{c})$ follows from the fact that
\begin{eqnarray*}
\left\{  W_{\bm{x}_i}^n - e^\alpha M_n \hat{W}_p^n \ge 0\right\} \subset 
\left\{ \sum_{j=1}^{M_n} W_{\bm{x}_j}^n - e^\alpha M_n \hat{W}_p^n \ge 0 \right\}
\end{eqnarray*}
holds for each $i$.

Now, we evaluate each term of \eqref{eq:proof-key-lemma-1} separately. 
Since $\bm{x}_i \in {\cal A}_n(\delta)$ for $i \in {\cal B}_n$, from Lemma \ref{lemma:output-of-non-typical}, 
we have $\hat{W}_p^n(\bm{y}) = 0$ for $\bm{y} \in T_{W,\delta}(\bm{x}_i)$, which implies 
\begin{eqnarray*}
T_{W,\delta}(\bm{x}_i) \subset \left\{  W_{\bm{x}_i}^n - e^\alpha M_n \hat{W}_p^n \ge 0\right\}.
\end{eqnarray*}
Thus, the first term is lower bounded as 
\begin{eqnarray}
\lefteqn{ \sum_{i \in {\cal B}_n} \frac{1}{M_n} W_{\bm{x}_i}^n \left\{  W_{\bm{x}_i}^n - e^\alpha M_n \hat{W}_p^n \ge 0\right\} } \nonumber \\
 &\ge& \sum_{i \in {\cal B}_n} \frac{1}{M_n} W_{\bm{x}_i}^n(T_{W,\delta}(\bm{x}_i)) \nonumber \\
 &\ge& \frac{|{\cal B}_n|}{M_n} (1-\gamma_n)
 \label{eq:proof-key-lemma-2}
\end{eqnarray}
for sufficiently large $n$. On the other hand, since 
\begin{eqnarray*}
\left\{  W_{\bm{x}_i}^n - e^\alpha M_n W_p^n \ge 0\right\} 
  \subset  \left\{  W_{\bm{x}_i}^n - e^\alpha M_n \hat{W}_p^n \ge 0\right\},
\end{eqnarray*}
the second term is lower bounded as 
\begin{eqnarray}
\lefteqn{ 
\sum_{i \in {\cal B}_n^c} \frac{1}{M_n} W_{\bm{x}_i}^n \left\{  W_{\bm{x}_i}^n - e^\alpha M_n \hat{W}_p^n \ge 0\right\} } \nonumber \\
&\ge& \sum_{i \in {\cal B}_n^c} \frac{1}{M_n} W_{\bm{x}_i}^n \left\{  W_{\bm{x}_i}^n - e^\alpha M_n W_p^n \ge 0\right\}.
\label{eq:proof-key-lemma-3}
\end{eqnarray}
Finally, by combining \eqref{eq:proof-key-lemma-0}-\eqref{eq:proof-key-lemma-3}, we have the desired bound.
\qed

\subsection{Proof of Lemma \ref{lemma:perturbation-expectation}}

Let 
\begin{eqnarray} \label{eq:definition-q-delta}
{\cal Q}(\delta) := \left\{ q : |W_q(b) - W_p(b) | \le 2 |{\cal X}| \delta~\forall b \in {\cal Y} \right\}.
\end{eqnarray}
Then, from the definition of ${\cal A}_n(\delta)$, we have
\begin{eqnarray} \label{eq:proof-lemma-perturbation-1}
\sum_a P_{\bm{x}}(a) D(W_a \| W_p) \ge \min_{q \in {\cal Q}(\delta)} \sum_a q(a) D(W_a \| W_p).
\end{eqnarray}
Since the righthand side of \eqref{eq:proof-lemma-perturbation-1} is a linear programming, by 
the perturbation analysis \cite[Sec.~5.6.2]{boyd-book:04}, we have
\begin{eqnarray*}
\lefteqn{  \min_{q \in {\cal Q}(\delta)} \sum_a q(a) D(W_a \| W_p)  } \\
&\ge&  \min_{q \in {\cal Q}(0)} \sum_a q(a) D(W_a \| W_p) - \tau(\delta) \\
&=& S_{W_p} - \tau(\delta)
\end{eqnarray*}
for some $\tau(\delta)$ such that $\tau(\delta) \to 0$ as $\delta \to 0$.
\qed

\subsection{Proof of Lemma \ref{lemma:perturbation-variance}}

Since \eqref{eq:condition} holds, we have ${\cal V}(p,W) = {\cal Q}(0)$, where 
${\cal Q}(\delta)$ is defined by \eqref{eq:definition-q-delta}. Thus, we have
\begin{eqnarray*}
V_{p,W}^- = \min_{q \in {\cal Q}(0)} V_{q,W_p},~~~~
V_{p,W}^+ = \max_{q \in {\cal Q}(0)} V_{q,W_p}.
\end{eqnarray*}
We also have
\begin{eqnarray} \label{eq:perturbation-variance-proof-1}
V_{P_{\bm{x}}, W_p} &\ge& \min_{q \in {\cal Q}(\delta)} V_{q,W_p}, \\
V_{P_{\bm{x}},W_p} &\le& \max_{q \in {\cal Q}(\delta)} V_{q,W_p}
 \label{eq:perturbation-variance-proof-2}
\end{eqnarray}
for $\bm{x} \notin {\cal A}_n(\delta)$. Since the righthand sides of \eqref{eq:perturbation-variance-proof-1}
and \eqref{eq:perturbation-variance-proof-2} are linear programmings, we can show the statement of
the lemma in the same reason as Lemma \ref{lemma:perturbation-expectation}.
\qed

\bibliographystyle{../../09-04-17-bibtex/IEEEtran}
\bibliography{../../09-04-17-bibtex/reference.bib}
\end{document}